\documentclass[aps,prl,preprint,tightenlines,superscriptaddress,showpacs,byrevtex]{revtex4}
\newcommand{\nbrlcpiapp}{$2.1\pm0.2\pm0.3\pm0.5$} 
\newcommand{\nbrlcpiazr}{$1.4\pm0.2\pm0.2\pm0.4$} 
\newcommand{\nbrlcpibpp}{$1.2\pm0.1\pm0.2\pm0.3$} 
 
\newcommand{\nbrnonr}{$6.4\pm0.4\pm0.9\pm1.7$}
\newcommand{\nbrtotal}{$11.2\pm0.5\pm1.4\pm2.9$}
\usepackage{graphicx} 
\usepackage{dcolumn}  

\graphicspath{{ps}}

\begin{document}

\preprint{\vbox{ 
\hbox{   }
                 \hbox{Belle Preprint 2006-25}
                 \hbox{KEK   Preprint 2006-36}
                 \hbox{HEP-EX/0608025}
                 \hbox{BELLE-CONF-0604}
}}

\title{ \quad\\[0.5cm] 
Study of the charmed baryonic decays \\ 
$\bar{B}^0\rightarrow\Sigma_c^{++}\bar{p}\pi^-$
and $\bar{B}^0\rightarrow\Sigma_c^{0}\bar{p}\pi^+$
}
\affiliation{Budker Institute of Nuclear Physics, Novosibirsk}
\affiliation{Chonnam National University, Kwangju}
\affiliation{University of Cincinnati, Cincinnati, Ohio 45221}
\affiliation{The Graduate University for Advanced Studies, Hayama, Japan} 
\affiliation{University of Hawaii, Honolulu, Hawaii 96822}
\affiliation{High Energy Accelerator Research Organization (KEK), Tsukuba}
\affiliation{Institute of High Energy Physics, Chinese Academy of Sciences, Beijing}
\affiliation{Institute of High Energy Physics, Vienna}
\affiliation{Institute of High Energy Physics, Protvino}
\affiliation{Institute for Theoretical and Experimental Physics, Moscow}
\affiliation{J. Stefan Institute, Ljubljana}
\affiliation{Kanagawa University, Yokohama}
\affiliation{Korea University, Seoul}
\affiliation{Swiss Federal Institute of Technology of Lausanne, EPFL, Lausanne}
\affiliation{University of Ljubljana, Ljubljana}
\affiliation{University of Maribor, Maribor}
\affiliation{University of Melbourne, Victoria}
\affiliation{Nagoya University, Nagoya}
\affiliation{Nara Women's University, Nara}
\affiliation{National Central University, Chung-li}
\affiliation{National United University, Miao Li}
\affiliation{Department of Physics, National Taiwan University, Taipei}
\affiliation{H. Niewodniczanski Institute of Nuclear Physics, Krakow}
\affiliation{Nippon Dental University, Niigata}
\affiliation{Niigata University, Niigata}
\affiliation{University of Nova Gorica, Nova Gorica}
\affiliation{Osaka City University, Osaka}
\affiliation{Osaka University, Osaka}
\affiliation{Panjab University, Chandigarh}
\affiliation{Peking University, Beijing}
\affiliation{RIKEN BNL Research Center, Upton, New York 11973}
\affiliation{University of Science and Technology of China, Hefei}
\affiliation{Seoul National University, Seoul}
\affiliation{Shinshu University, Nagano}
\affiliation{Sungkyunkwan University, Suwon}
\affiliation{University of Sydney, Sydney NSW}
\affiliation{Tata Institute of Fundamental Research, Bombay}
\affiliation{Toho University, Funabashi}
\affiliation{Tohoku Gakuin University, Tagajo}
\affiliation{Tohoku University, Sendai}
\affiliation{Department of Physics, University of Tokyo, Tokyo}
\affiliation{Tokyo Institute of Technology, Tokyo}
\affiliation{Tokyo Metropolitan University, Tokyo}
\affiliation{Tokyo University of Agriculture and Technology, Tokyo}
\affiliation{Virginia Polytechnic Institute and State University, Blacksburg, Virginia 24061}
\affiliation{Yonsei University, Seoul}
  \author{K.~S.~Park}\affiliation{Sungkyunkwan University, Suwon} 
  \author{H.~Kichimi}\affiliation{High Energy Accelerator Research Organization (KEK), Tsukuba} 
  \author{K.~Abe}\affiliation{High Energy Accelerator Research Organization (KEK), Tsukuba} 
  \author{K.~Abe}\affiliation{Tohoku Gakuin University, Tagajo} 
  \author{I.~Adachi}\affiliation{High Energy Accelerator Research Organization (KEK), Tsukuba} 
  \author{H.~Aihara}\affiliation{Department of Physics, University of Tokyo, Tokyo} 
  \author{D.~Anipko}\affiliation{Budker Institute of Nuclear Physics, Novosibirsk} 
  \author{K.~Arinstein}\affiliation{Budker Institute of Nuclear Physics, Novosibirsk} 
  \author{V.~Aulchenko}\affiliation{Budker Institute of Nuclear Physics, Novosibirsk} 
  \author{T.~Aushev}\affiliation{Swiss Federal Institute of Technology of Lausanne, EPFL, Lausanne}\affiliation{Institute for Theoretical and Experimental Physics, Moscow} 
% ...
\collaboration{The Belle Collaboration}

\date{\today}

\begin{abstract}
We study the three-body charmed baryonic decays 
$\bar{B}^0\rightarrow\Sigma_{ {c}}^{++}\bar{p}\pi^-$
and $\bar{B}^0\rightarrow\Sigma_{ {c}}^{0}\bar{p}\pi^+$ 
in the four-body final state $\bar{B}^0\rightarrow\Lambda_{ {c}}^{+}\bar{p}\pi^+\pi^-$,
using a data sample of 357\,fb$^{-1}$
accumulated at the $\Upsilon(4S)$ resonance with the Belle detector at the KEKB asymmetric-energy $e^+e^-$ collider.  
We measure the branching fractions
${\cal B} (\bar{B}^0\rightarrow\Sigma_{ {c}}(2455)^{++}\bar{p}\pi^-)$=(\nbrlcpiapp)$\times 10^{-4}$,
${\cal B}(\bar{B}^0\rightarrow\Sigma_{ {c}}(2455)^{0}\bar{p}\pi^+)$=(\nbrlcpiazr)$\times10^{-4}$ and
${\cal B}(\bar{B}^0\rightarrow\Sigma_{ {c}}(2520)^{++}\bar{p}\pi^-)$=(\nbrlcpibpp)$\times 10^{-4}$ 
with signal significances of 13.1$\,\sigma$, 9.4$\,\sigma$ and 7.1$\,\sigma$, respectively.
The errors are statistical, systematic, and due to the uncertainty in
${\cal B}(\Lambda_{ {c}}^+\rightarrow{p}K^-\pi^+)$, respectively. 
We also set an upper limit
${\cal B}(\bar{B}^0\rightarrow\Sigma_{ {c}}(2520)^{0}\bar{p}\pi^+)<0.38\times10^{-4}$ at the 90\% confidence level. 
In addition, we obtain a non-resonant branching fraction of 
(\nbrnonr)$\times10^{-4}$, and a total branching fraction of
(\nbrtotal)$\times10^{-4}$ for $\bar{B}^0\rightarrow\Lambda_{ {c}}^{+}\bar{p}\pi^+\pi^-$.
\end{abstract}
\pacs{ 13.25.Hw, 14.20.Lq, 14.40.Nd }  

\maketitle

\normalsize

\newpage

\normalsize

\vskip 1.0cm

The large mass of the $b$-quark enables $B$ mesons to decay into two baryons with additional pions. 
Since the CKM matrix element $|V_{cb}|$~\cite{ckm} is substantially larger than $|V_{ub}|$, 
these baryonic decays preferentially proceed through $b\rightarrow c$ transitions and produce final states rich 
in charmed baryons. 
CLEO pioneered the study of these processes and reported branching fractions and evidence of 
several exclusive charmed baryonic decays with a 9.1 fb$^{-1}$ data sample~\cite{cleo-lamc,cleo-ppbar,cleo_blamc}.
Recently, Belle has observed several new decay modes into two-, 
three-, and four-body final states with charmed baryons~\cite{belle_blamc,belle_lamc1},
and three-body decays with charmless baryons~\cite{belle_ppk,belle_Dpp,belle_plambdapi,belle_pph,belle_jpsi_lam_pbar}.
We find a hierarchy of the branching fractions that depends on the multiplicity in the final state:
$\sim 2\times10^{-5}$ for the two-body decays $\bar{B}^0\rightarrow\Lambda_c^+\bar{p}$ and
$B^-\rightarrow \Sigma_c^0(2455/2520)\bar{p}$, 
$\sim 1\times10^{-4}$ for the three-body decay $B^-\rightarrow \Lambda_c^+\bar{p}\pi^-$, 
and $\sim 7\times10^{-4}$ for the four-body decay $\bar{B}^0\rightarrow \Lambda_c^+\bar{p}\pi^+\pi^-$~\cite{belle_lamc1,beach04}.
There are several theoretical models that describe the decay mechanisms and predict
the branching fractions of baryonic $B$ decays into two-body and three-body final states
~\cite{jarfi,chernyak,theory_charmless,theory_ppk,theory_Dpp,theory_plambdapi,theory_rosner,lamc2_last_theory}.
Detailed studies of such decays 
are very important to provide strict constraints on these theoretical models.

In this paper, we report improved measurements of the intermediate three-body decays 
$\bar{B}^0\rightarrow\Sigma_{ {c}}(2455/2520)^{++}\bar{p}\pi^-$
and $\bar{B}^0\rightarrow\Sigma_{ {c}}(2455/2520)^{0}\bar{p}\pi^+$
in the four-body final state $\bar{B}^0\rightarrow\Lambda_{ {c}}^{+}\bar{p}\pi^+\pi^-$.
This study is based on a 357\,fb$^{-1}$ data sample
accumulated at the $\Upsilon(4S)$ with the Belle detector at the KEKB asymmetric-energy
$e^+ e^-$ collider~\cite{kekb}. 

The Belle detector is a large-solid-angle spectrometer based on 
a 1.5~Tesla superconducting solenoid magnet. It consists of a silicon vertex detector (a three-layer 
silicon vertex detector (SVDI) for the first sample of $(152.0 \pm 1.2)\times10^6$ $B\bar{B}$ events
and a four-layer silicon vertex detector (SVDII) for the latter $(235.8 \pm 3.6)\times10^6$ $B\bar{B}$ events),
a 50-layer central drift chamber(CDC), an array of aerogel threshold Cherenkov counters (ACC),
a barrel-like arrangement of time-of-flight scintillation counters (TOF), and an electromagnetic
calorimeter comprised of CsI\,(Tl) crystals located inside the superconducting 
solenoid coil.
An iron flux return located outside the coil is instrumented to detect $K_L^0$ mesons
and to identify muons. The detector is described in detail elsewhere~\cite{belle}. 

We simulate the detector response and estimate the efficiency for signal reconstruction by Monte Carlo simulation (MC). 
We use the QQ program~\cite{qq98} for signal event generation and
a GEANT-based detector simulation program~\cite{geant}.
A sample of $5.45\times10^4$ signal events is generated for each of the
four-body decay
$\bar{B}^0\rightarrow\Lambda_{ {c}}^+\bar{p}\pi^+\pi^-$,
the intermediate three-body decays
$\bar{B}^0\rightarrow\Sigma_{ {c}}(2455/2520)^{++}\bar{p}\pi^-$ and
$\bar{B}^0\rightarrow\Sigma_{ {c}}(2455/2520)^0\bar{p}\pi^+$, and
their charge conjugate modes. Each signal sample is
processed by the detector simulation program that takes into account the 
differences between SVDI and SVDII as well as the long-term variation of 
the beam background conditions.
 
The mode $\bar{B^0}\rightarrow\Lambda_{ {c}}^+ \bar{p} \pi^-\pi^+$
is tagged by an associated $\Lambda_{ {c}}^+$ particle, which decays
into $p K^-\pi^+$. Charge-conjugate modes 
are implicitly included throughout this paper unless noted otherwise.
To reconstruct $\Lambda_{ {c}}^+$ and $\bar{B^0}$ signals, we require tracks 
to have distances-of-closest-approach to the interaction point of less 
than 5.0\,cm in $z$ (the direction opposite to the $e^+$ beam direction) and 1.0\,cm in a plane perpendicular to the $z$-axis.
We require the $\Lambda_{ {c}}^+$ mass to be within $\pm0.014{\,\rm GeV/c^2}$ ($\sim3.5\sigma$) of our fitted mass of 
$2.287\,\rm{GeV/c^2}$.
Hadrons such as protons, kaons and pions are identified by using likelihood ratios provided  
from the CDC $dE/dx$, TOF and ACC information (PID)~\cite{belle-pid}. 
We use likelihood ratios $L_{s}/(L_{s}+L_{b})$, where $s$ and
$b$ stand for the hadron species to be identified and for the background, 
respectively. We require these ratios to be greater than 0.6, 0.6
and 0.4 for proton, kaon and pion selection, respectively.
The efficiency for proton identification is 95\% with a kaon fake rate of 1.0\%.
The efficiencies for kaons and pions are about 90\%;
the corresponding pion and kaon misidentification rates are about 10\%~\cite{belle-pid}.
Tracks that are positively identified as electrons or muons are rejected.
We impose loose requirements on the vertex fit $\chi^2$'s for the tracks 
from $\Lambda_{ {c}}^+\rightarrow p K^-\pi^+$ ($\chi^2_{\Lambda_c^+}$) and
$\bar{B}^0\to\Lambda_c^+\bar{p}\pi^+\pi^-$ ($\chi^2_{B}$)
to reject background from the decay products of $K_S^0$ 
and $\Lambda$ particles.
When there are multiple $B$ candidates in an event,
we choose the candidate with the smallest $\chi^2_{B}$.

To suppress the continuum background ($u,d,s,c$ pair production), 
we use two event 
shape variables, $R_2$ and
$\cos\theta_{ \mathrm{thrust}}$. The variable $R_2$ is the ratio of the second to zeroth
order Fox-Wolfram moments~\cite{fox_wolfram}, and $\cos\theta_{\mathrm{thrust}}$ is defined 
as the cosine of the angle between the thrust axis of the reconstructed 
$B$ decay products and the thrust axis of the other tracks
in the center-of-mass system(CMS).
We require $R_2\le0.35$ and $|\cos\theta_{\mathrm{thrust}}|\le0.8$, which 
retain 80\% of the signal and remove 60\% of the continuum background.
These requirements, together with the final $B$ signal selection discussed below, 
reduce the continuum background by a factor of about $10^{4}$.

The final selection requirements are based on the kinematic variables $\Delta{E}$ and 
$M_{\mathrm{bc}}$. The variable $\Delta{E}=E_{B}-E_{\mathrm{beam}}$ is 
the difference between the reconstructed $B$ meson energy ($E_{B}$) and
the beam energy ($E_{\mathrm{beam}}$) evaluated in the CMS, 
while $M_\mathrm{bc}=\sqrt{E_{\mathrm{beam}}^2 - P_B^2}$
is the beam energy constrained $B$ meson mass. Here,
${P_B}$ is the momentum of the $B$ meson also evaluated in the CMS.

\begin{figure}[!htb]
\centering
\includegraphics[width=0.45\textwidth]{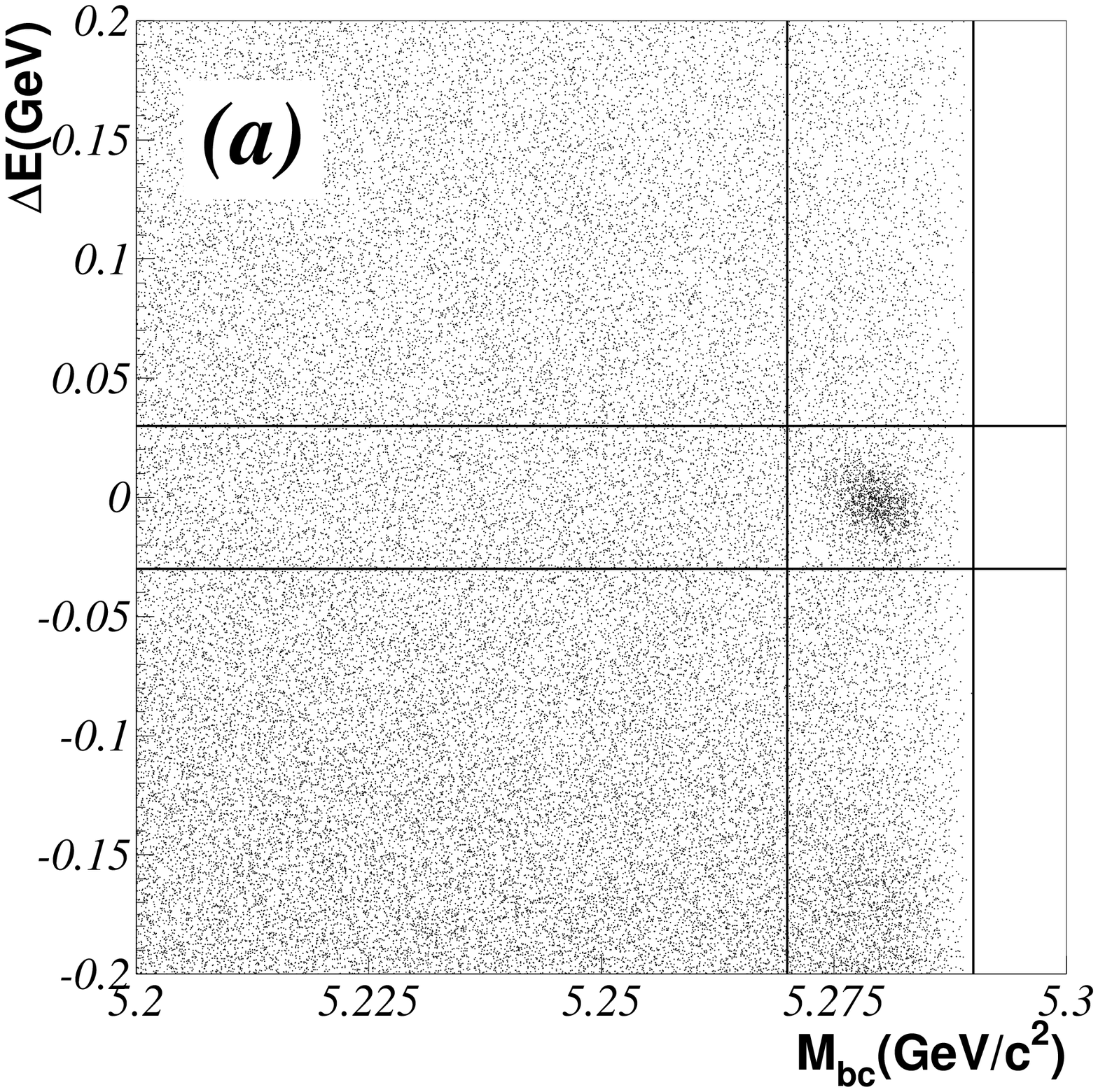}
\includegraphics[width=0.45\textwidth]{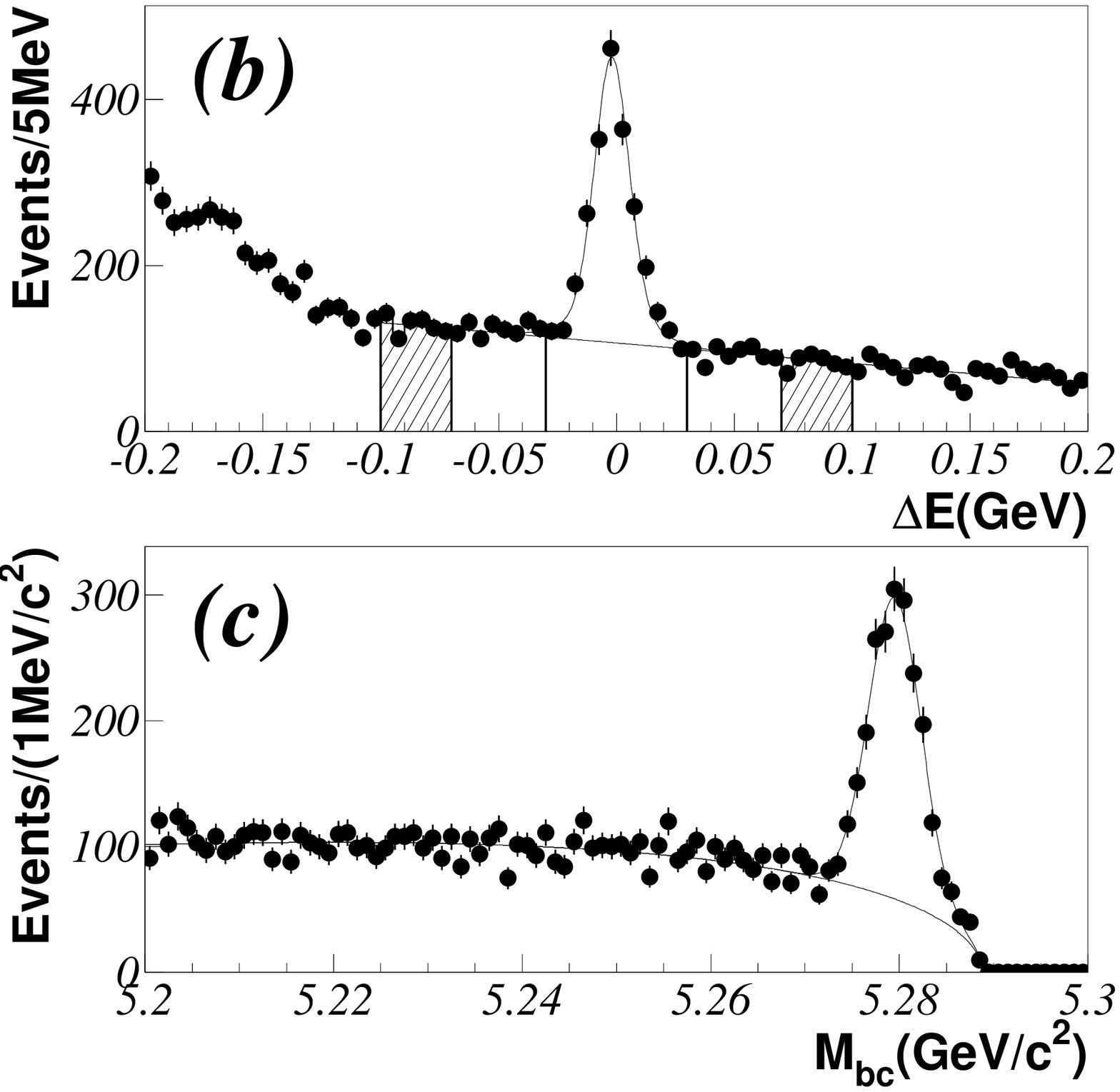}
\centering
\caption{
(a) Scatter plot of $\Delta{E}$ vs.\ $M_{\mathrm{bc}}$ for 
$\bar{B}^{0}\rightarrow\Lambda_{{c}}^+ \bar{p}\pi^+\pi^-$ signal candidates. 
(b) The $\Delta{E}$ distribution for $M_{\mathrm{bc}}\ge5.27\,{\rm GeV}/c^2$.
The shaded regions indicate the sideband whose total area is equal
to the background in the $B$ signal region.
(c) The $M_{\mathrm{bc}}$ distribution for $|\Delta{E}|\le0.03\,{\rm GeV}$. 
}
\label{fig:2d_de_mbc}
\end{figure}

Figure~\ref{fig:2d_de_mbc} (a) shows a scatter plot of $\Delta{E}$ vs.\ $M_{\mathrm{bc}}$.
The vertical lines show the $B$ signal region of
$5.27\,{\rm GeV}/c^2\le M_{\mathrm{bc}}\le5.29\,{\rm GeV}/c^2$,
and the horizontal lines indicate the signal region, $|\Delta{E}|\le0.03$\,GeV.
Figure~\ref{fig:2d_de_mbc} (b) shows the $\Delta{E}$ distribution for 
the $M_{\mathrm{bc}}$ signal region,
where the curve shows the result of the fit with a double Gaussian for the signal and a linear background 
in the fit interval of $-0.1$\,GeV$\le\Delta{E}\le0.2$\,GeV.
Figure~\ref{fig:2d_de_mbc} (c) is the $M_{\mathrm{bc}}$ distribution for $|\Delta{E}|\le0.03$\,GeV.
The curve shows the fit with a single Gaussian for the signal and an ARGUS function~\cite{argus_function} for the background.
We use the $\Delta{E}$ distribution to determine the $B$ signal yield, as we find 
a peaking background in the $M_{\mathrm{bc}}$ distribution from a study of the $\Delta{E}$ sideband.

To remove the feed-down from higher multiplicity modes with additional pions, 
we restrict the fit region to $\Delta{E}\ge-0.1$\,GeV.
The signal shape parameters are fixed to those fitted to the 
corresponding MC, where
we find the Gaussian widths $\sigma_1$ (with a ratio of $\sigma_2/\sigma_1$) of
$6.6\pm0.4\,\rm{MeV}/c^2$ ($2.2\pm0.2$) for SVDI, and 
$7.2\pm0.2\,\rm{MeV}/c^2$ ($2.3\pm0.2$) for SVDII.
We obtain $B$ signals of $535\pm30$ and $865\pm38$ events for SVDI and SVDII data, 
respectively.
The efficiency-corrected signal yields normalized to the number of the $B\bar{B}$ events are consistent within errors.
Thus, we combine the SVDI and SVDII data and obtain the total $\bar{B}^0\to\Lambda_c^+\bar{p}\pi^+\pi^-$ yield of $1400\pm49$ events. 

\begin{figure}[!htb]
\includegraphics[width=0.80\textwidth]{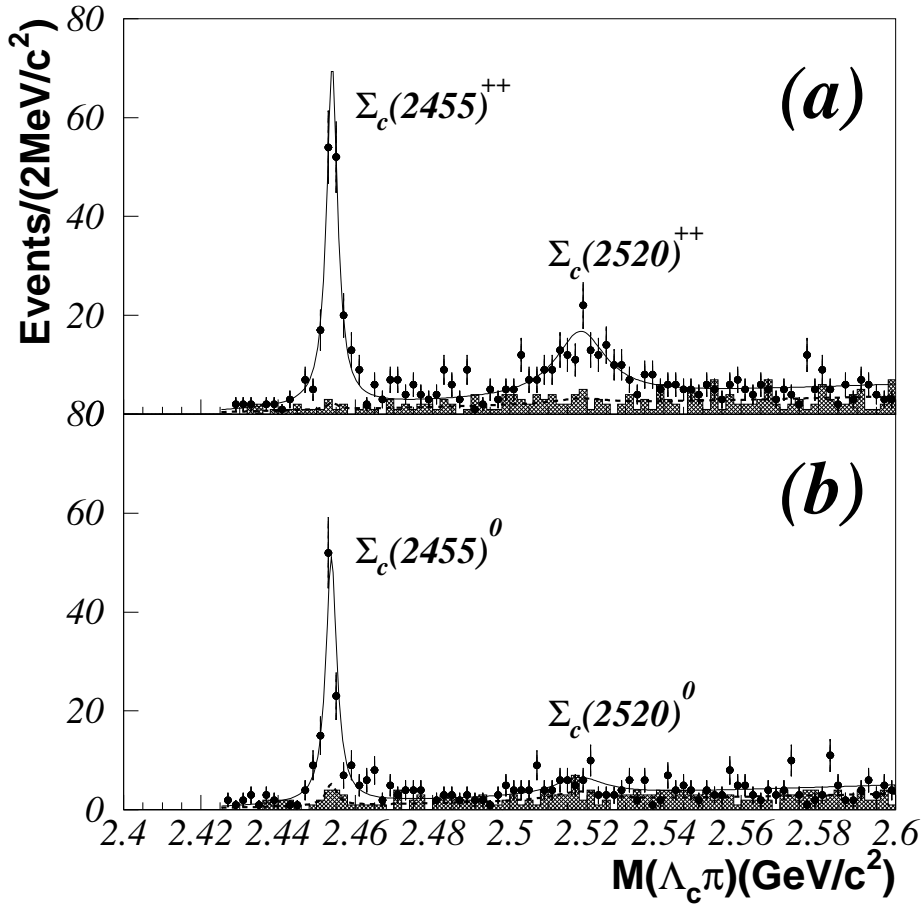}
\caption{
The mass distributions of (a) $\Lambda_{ {c}}^+\pi^+$ and 
(b) $\Lambda_{ {c}}^+\pi^-$ for  $\bar{B}^0\rightarrow\Lambda_{ {c}}^+\bar{p}\pi^+\pi^-$.
The points with error bars show the mass distribution for the events in the $B$ signal region, and
the shaded histogram indicates that for the sideband region.
See the text for details.
}
\label{fig:sig_mass_full_win1}
\end{figure}

Figure~\ref{fig:sig_mass_full_win1} shows (a) the $\Lambda_{ {c}}^+\pi^+$ and (b) the 
$\Lambda_{ {c}}^+\pi^-$ mass distributions 
for $\bar{B}^0\rightarrow\Lambda_{ {c}}^+\bar{p}\pi^+\pi^-$.
We observe clear peaks for the $\Sigma_{ {c}}(2455)^{++/0}$ and $\Sigma_{ {c}}(2520)^{++}$.
The points with error bars show the events in the $B$ signal region defined by 
$|\Delta{E}|\le0.03\,\rm{GeV}$ and $5.27\,{\rm GeV}/c^2\le M_{\mathrm{bc}}\le5.29\,{\rm GeV}/c^2$.
The shaded histograms are the events in the $\Delta{E}$ 
sideband region defined by $0.07\,{\rm GeV}<|\Delta E|<0.10\,\rm{GeV}$ 
and $5.27\,{\rm GeV}/c^2\le M_{\mathrm{bc}}\le5.29\,{\rm GeV}/c^2$. 
As seen in Figure~\ref{fig:2d_de_mbc} (b), the number of the sideband events is equal to that of the background 
in the $B$ signal region of $|\Delta{E}|\le0.03\,\rm{GeV}$.
We find small peaking backgrounds near the 
$\Sigma_c(2455/2520)^0$ masses in the sideband events.

To obtain the $\Sigma_c(2455)^{++/0}$ and $\Sigma_c(2520)^{++/0}$ signal yields,
we consider possible contributions from peaking backgrounds seen in the sideband.
We also study the background shape using MC samples for the non-resonant four-body decay
$\bar{B}^0\to\Lambda_{ {c}}^{+}\bar{p}\pi^+\pi^-$, and
the intermediate three-body decays 
$\bar{B}^0\to\Sigma_{ {c}}(2455/2520)^{++}\bar{p}\pi^-$ and
$\bar{B}^0\to\Sigma_{ {c}}(2455/2520)^{0}\bar{p}\pi^+$. 
We find a linear behavior for
the mass distributions from non-resonant $\Lambda_{ {c}}^+\pi^{\pm}$ combinations. 
Therefore, we introduce independent linear background functions in the $B$ signal and
the sideband regions.
 
We perform a simultaneous binned likelihood fit to the mass distributions 
with the following functions 
\begin{equation}
(N_{1s}+N_{1b})\times BW_1(M) + (N_{2s} + N_{2b})\times BW_2(M) + (c_s + a_s\times{M})
\end{equation} 
\begin{equation}
N_{1b}\times BW_1(M) + N_{2b}\times BW_2(M) + (c_b + a_b\times{M}) 
\end{equation} 
for the $B$ signal and the sideband events, respectively.
Here, $M$ is the $\Lambda_{ {c}}^+\pi^{\pm}$ mass, and
$BW(M)$ represents a Breit-Wigner function.
The subscripts 1 and 2 indicate $\Sigma_{ {c}}(2455)$ and $\Sigma_{ {c}}(2520)$, 
and the subscripts $s$ and $b$ stand for the signal and sideband, respectively. 
$c$ and $a$ are parameters of the linear functions. 
$N_{1s}$ and $N_{2s}$ are the net signal yields of $\Sigma_{ {c}}(2455)$ and $\Sigma_{ {c}}(2520)$,
respectively, and $N_{1b}$ and $N_{2b}$ are the normalizations of the peaking backgrounds in the sidebands.

Table~\ref{table:mass-width} lists the parameters 
$M_{\mathrm{PDG}}$ and $\Gamma_\mathrm{PDG}^{\mathrm{eff}}$ for the Breit-Wigner functions $BW_1$ and $BW_2$ 
used in the simultaneous fit. 
When we float these parameters, the fitter obtains values
$M_\mathrm{fit}$ and $\Gamma_\mathrm{fit}^{\mathrm{eff}}$, consistent with the
PDG values $M_\mathrm{PDG}$ and $\Gamma_\mathrm{PDG}^{\mathrm{eff}}$~\cite{pdg2006}
(the latter is $\Gamma_\mathrm{PDG}$
convoluted with the Belle detector resolution). 
Thus, we fix those parameters to the PDG values; the fitted parameters in the fit are then
the signal yields $N_{1s}$ and $N_{2s}$, the peaking background yields $N_{1b}$ and $N_{2b}$, 
and the linear background shape parameters $c_s$, $a_s$, $c_b$ and $a_b$.
The uncertainties in the signal yields due to the assumed masses and widths
are taken into account in systematic errors as discussed below.

\begin{table*}
\caption{ 
The Breit-Wigner parameters of $\Sigma_c(2455/2520)$ resonances. 
In the simultaneous fits, they are fixed to
$M_\mathrm{PDG}$ and $\Gamma_\mathrm{PDG}^{\mathrm{eff}}$.
$M_\mathrm{fit}$ and $\Gamma_\mathrm{fit}^{\mathrm{eff}}$ are the fitted values
to the data with statistical errors only. 
}
\begin{center}
\begin{tabular}{ccccc}  \hline 
 Resonances  & $M_\mathrm{fit}$(MeV$/c^2$) & $\Gamma_\mathrm{fit}^{\mathrm{eff}}$ (MeV$/c^2$) & 
$M_\mathrm{PDG}$(MeV$/c^2$) & $\Gamma_\mathrm{PDG}^{\mathrm{eff}}$(MeV$/c^2$) \\ \hline
$\Sigma_{ {c}}(2455)^{++}$ & $2454.1\pm0.2$ & $3.5\pm0.5$ & $2454.0\pm0.2$ & $3.44\pm0.30$ \\
$\Sigma_{ {c}}(2455)^{0}$  & $2453.4\pm0.6$ & $3.4\pm0.4$ & $2453.8\pm0.2$ & $3.44\pm0.40$ \\
$\Sigma_{ {c}}(2520)^{++}$ & $2517.9\pm1.4$ & $19.9\pm3.5$ & $2518.4\pm0.6$ &$16.0\pm2.0$ \\
$\Sigma_{ {c}}(2520)^{0}$  & $2514.3\pm2.8$ & $19.1\pm5.7$ & $2518.0\pm0.5$ &$16.0\pm2.0$ \\  \hline
\end{tabular}
\end{center}
\label{table:mass-width}
\end{table*}

In the fits shown in Figure~\ref{fig:sig_mass_full_win1},
we obtain $\chi^2$/n.d.f = 183.4/192 and 196.6/192 for the fits to
$\Lambda_{ {c}}^+\pi^+$ and $\Lambda_{ {c}}^+\pi^-$ mass distributions, respectively.
The solid curves show the fits to the mass distributions for the $B$ signal region,
and the dashed curves indicate the fits for the sideband region.
The significance of the $\Sigma_{ {c}}(2455)$\,($\Sigma_{ {c}}(2520)$) signal
is evaluated as $S=\sqrt{-2\ln(L_0 / L_{\mathrm{max}})}$, where 
$L_{\mathrm{max}}$ is the maximum likelihood of the fit and
$L_0$ is the likelihood for a fit with the yield 
of $\Sigma_{ {c}}(2455)$\,($\Sigma_{ {c}}(2520)$) fixed to zero
and the other parameters floated.
We study the change in the signal significances by varying the fixed masses and widths by their $\pm1\sigma$
errors and find that the resulting change is negligibly small. 

Table~\ref{table:br-final} summarizes the fitted signal yields, efficiencies,
significances and the branching fractions
obtained for intermediate three-body decays 
$\bar{B}^0\to\Sigma_{ {c}}(2455)^{++(0)}\bar{p}\pi^{-(+)}$ and
$\bar{B}^0\to\Sigma_{ {c}}(2520)^{++(0)}\bar{p}\pi^{-(+)}$. 
The third error is due to the uncertainty in the branching fraction of
${\cal B}(\Lambda_{ {c}}^+\rightarrow{p}K^-\pi^+)=(5.0\pm1.3)\%$. 
As a check, we calculate separate branching fractions
for charge-conjugate modes; the two branching fractions are in good agreement.

We obtain a systematic error of 11.7\% as a quadratic sum of 7.2\% due to track reconstruction efficiency,
9.1\% from the PID (both are coherent sums over the six tracks for the $B$ decay products)
and 1.9\% due to the uncertainty on $N(B\bar{B})$ and limited MC statistics. 
These errors are common to all decay modes.
The signal efficiencies in Table~\ref{table:br-final}
include the MC PID correction factor of $0.867\pm0.079$,
to account for a systematic difference between data and MC. 
Separate PID correction factors for proton, kaon and pion tracks 
as functions of momentum and azimuthal angle are determined from a comparison 
of data and MC for large samples of $D^{*+}\rightarrow D^0(K\pi)\pi^{+}$ 
and $\Lambda\rightarrow p \pi^-$ decays. The overall PID correction factor is then calculated
as a coherent sum over the six tracks for the selected $B$ signal events.
The error of $\pm0.079$ is taken into account as the PID systematic error of 9.1\% as mentioned above.
We estimate an error of 3.5\% for the total $B$ signal yield 
from the maximum variation of the yield in fits to the $\Delta{E}$ distribution
with the double Gaussian fixed to MC and with the shape paramters floated.
This uncertainty in the $B$ signal yield 
results in an error of 5.3\% for the signal yield of the non-resonant four-body decay (see below).
We estimate an error of 4.8\%\,(9.1\%) for $\Sigma_{ {c}}(2455)^{++/0}$\,($\Sigma_{ {c}}(2520)^{++/0}$)
from the variation in the fitted signal yield 
due to a $\pm 1\sigma$ change (0.4\,(2.0)\,MeV/$c^2$) in the width $\Gamma_\mathrm{PDG}^{\mathrm{eff}}$.
We find a negligibly small effect on the mass.
In addition, we take into account the uncertainty in the signal efficiency due to 
differences between the resonant substructure in data and signal MC.
The $\Sigma_{ {c}}(2455)^{++}$ data is consistent with three-body MC,
while the $\Sigma_{ {c}}(2455)^{0}$ data shows a broad $\bar{p}\pi^+$ mass structure  
that differs from MC phase space.
We estimate an error of 4.6\% for the $\Sigma_{ {c}}(2455)^{0}$ efficiency. 

We investigate the signal yield for non-resonant ${\bar{B}^0\rightarrow\Lambda_{ {c}}^+\bar{p}\pi^+\pi^-}$ decay,
which consists of four-body decay, as well as contributions
from decay modes with possible final state interactions or
resonance states of the $\bar{p}\pi^{\pm}$, $\pi^+\pi^-$ and $\Lambda_{ {c}}^+\bar{p}$ systems.
The signal efficiencies for $\bar{p}\pi^{\pm}$ tend to be lower than that for the non-resonant four-body $B$ decay
near the mass threshold.
We study two-body submass distributions 
and find some deviation from phase space near the threshold.
However, due to limited statistics we cannot draw any strong conclusions about possible resonant structures.
We conservatively estimate an uncertainty in the signal efficiency due to resonant structure to be 5\%.
Adding those errors in quadrature, we obtain total systematic errors of 
12.6\% for $\Sigma_c(2455)^{++}$, 
13.5\% for $\Sigma_c(2455)^{0}$, 
14.8\% for $\Sigma_c(2520)^{++/0}$,
and 13.7\% for non-resonant four-body $B$ decay. 

We obtain the branching fraction of the non-resonant four-body decay 
by subtracting the signal yields for the observed three-body decays from the total $B$ signal of $1400\pm49$ events
and correcting for the efficiency of non-resonant four-body MC. 
The total branching fraction 
is obtained by adding the branching fractions of the intermediate three-body and 
non-resonant four-body decay modes. 
The branching fractions are consistent with the previous 
measurements~\cite{cleo_blamc,belle_blamc} and
supersede our previous measurements~\cite{belle_blamc}. 

\begin{table*}
\caption{ 
Branching fractions for 
${\bar{B}^0\rightarrow\Sigma_{ {c}}(2455/2520)^{++}\bar{p}\pi^-}$ and
${\bar{B}^0\rightarrow\Sigma_{ {c}}(2455/2520)^{0} \bar{p} \pi^+}$.
The errors in the branching fractions are statistical, systematic
and due to the uncertainty in ${\cal B}(\Lambda_{ {c}}^+\rightarrow p K^-\pi^+)=5.0\pm1.3\%$,
respectively. See text for details of the systematic errors.
}
\begin{center}
\begin{tabular}{cccccc}  \hline 
   Modes   & Yield &  Det.eff.($\%$) & Sys.err(\%)
   &  Sign.($\sigma$) & ${\cal B}(\times 10^{-4})$ \\ \hline
${\bar{B}^0\rightarrow\Sigma_{ {c}}(2455)^{++}\bar{p}\pi^-}$ &
            $182\pm15$ &   4.57  & 12.6 & 13.1 &  \nbrlcpiapp \\
${\bar{B}^0\rightarrow\Sigma_{ {c}}(2455)^{0}\bar{p}\pi^+}$  &
           $122\pm14$  &   4.41  & 13.5 & 9.4 & \nbrlcpiazr \\ \hline
${\bar{B}^0\rightarrow\Sigma_{ {c}}(2520)^{++}\bar{p}\pi^-}$ &
           $155\pm18$ &    6.91  & 14.8 & 7.1 & \nbrlcpibpp \\ 
${\bar{B}^0\rightarrow\Sigma_{ {c}}(2520)^{0}\bar{p}\pi^+}$ &
          $22\pm16$ &  6.75 & 14.8 & 1.3  & $<0.38$ (90\%C.L.) \\ \hline
Non-resonant & $919\pm58$  &   7.50 &   13.7 &
                  & \nbrnonr  \\  \hline \hline
Total  &  $1400\pm49$  &     
                   & & & \nbrtotal \\ \hline
\end{tabular}
\end{center}
\label{table:br-final}
\end{table*}

In summary, we study the three-body charmed baryonic decays 
$\bar{B}^0\rightarrow\Sigma_{ {c}}^{++}\bar{p}\pi^-$
and $\bar{B}^0\rightarrow\Sigma_{ {c}}^{0}\bar{p}\pi^+$ 
in the four-body final state $\bar{B}^0\rightarrow\Lambda_{ {c}}^{+}\bar{p}\pi^+\pi^-$,
and measure the branching fractions
${\cal B}(\bar{B}^0\rightarrow\Sigma_{ {c}}(2455)^{++}\bar{p}\pi^-)$=(\nbrlcpiapp)$\times 10^{-4}$,
${\cal B}(\bar{B}^0\rightarrow\Sigma_{ {c}}(2455)^{0}\bar{p}\pi^+)$=(\nbrlcpiazr)$\times10^{-4}$ and
${\cal B}(\bar{B}^0\rightarrow\Sigma_{ {c}}(2520)^{++}\bar{p}\pi^-)$=(\nbrlcpibpp)$\times 10^{-4}$ 
with signal significances of 13.1$\,\sigma$, 9.4$\,\sigma$ and 7.1$\,\sigma$, respectively.
The errors are statistical, systematic, and due to the uncertainty in 
${\cal B}(\Lambda_{ {c}}^+\rightarrow{p}K^-\pi^+)$, respectively. 
We also set an upper limit
${\cal B}(\bar{B}^0\rightarrow\Sigma_{ {c}}(2520)^{0}\bar{p}\pi^+)<0.38\times10^{-4}$ at the 90\% confidence level.
In addition, we obtain a non-resonant branching fraction of 
(\nbrnonr)$\times10^{-4}$, and a total branching fraction of
(\nbrtotal)$\times10^{-4}$ for $\bar{B}^0\rightarrow\Lambda_{ {c}}^{+}\bar{p}\pi^+\pi^-$.

The observed branching fraction ${\cal B}(\bar{B}^0\rightarrow\Sigma_{ {c}}(2455)^{++}\bar{p}\pi^-)$ is
comparable to a previous measurement of ${\cal B}(B^-\rightarrow\Lambda_{ {c}}^+\bar{p}\pi^-)$ 
by Belle~\cite{hep-ex0409005}. 
The $\Sigma_{ {c}}(2455)^{++}$ mode has a larger branching fraction 
than the $\Sigma_{ {c}}(2455)^{0}$ and $\Sigma_{ {c}}(2520)^{++}$ modes, 
and the $\Sigma_{ {c}}(2520)^{0}$ mode is significantly suppressed.  
The branching fraction ${\cal B}(\bar{B}^0\rightarrow\Sigma_{ {c}}(2455/2520)^{++}\bar{p}\pi^-)$ is larger than 
${\cal B} (\bar{B}^0\rightarrow\Sigma_{ {c}}(2455/2520)^{0}\bar{p}\pi^+)$, probably due to an additional contribution
from an external $W$ emission diagram~\cite{lamc2_last_theory}. 
The total branching fraction for the four-body decay 
${\cal B}(\bar{B}^0\rightarrow\Lambda_{ {c}}^+\bar{p}\pi^+\pi^-)$
is five times larger than ${\cal B}(B^-\rightarrow\Lambda_{ {c}}^{+}\bar{p}\pi^-)$, as it consists of
both intermediate three-body decays $\bar{B}^0\to\Sigma_{ {c}}(2455/2520)\bar{p}\pi$ ($\sim$40\%) 
and a non-resonant four-body decay ($\sim$60\%).

\vspace {0.5cm}
ACKNOWLEDGEMENT

\vspace {0.5cm}
We thank the KEKB group for the excellent operation of the
accelerator, the KEK cryogenics group for the efficient
operation of the solenoid, and the KEK computer group and
the National Institute of Informatics for valuable computing
and Super-SINET network support. We acknowledge support from
the Ministry of Education, Culture, Sports, Science, and
Technology of Japan and the Japan Society for the Promotion
of Science; the Australian Research Council and the
Australian Department of Education, Science and Training;
the National Science Foundation of China and the Knowledge
Innovation Program of the Chinese Academy of Sciences under
contract No.~10575109 and IHEP-U-503; the Department of
Science and Technology of India; 
the BK21 program of the Ministry of Education of Korea, 
the CHEP SRC program and Basic Research program 
(grant No.~R01-2005-000-10089-0) of the Korea Science and
Engineering Foundation, and the Pure Basic Research Group 
program of the Korea Research Foundation; 
the Polish State Committee for Scientific Research; 
the Ministry of Science and Technology of the Russian
Federation; the Slovenian Research Agency;  the Swiss
National Science Foundation; the National Science Council
and the Ministry of Education of Taiwan; and the U.S.\
Department of Energy.


\begin{thebibliography}{99}
\bibitem{qq98}
  Events are generated with the CLEO group's QQ program.
\end{thebibliography}
\end{document}